\pdfoutput=1
\documentclass[11pt]{article}
\usepackage{EMNLP2023}
\usepackage{times}
\usepackage{latexsym}
\usepackage[T1]{fontenc}
\usepackage[utf8]{inputenc}
\usepackage{microtype}
\usepackage{inconsolata}
\usepackage{times}
\usepackage{latexsym}
\usepackage[T1]{fontenc}
\usepackage[utf8]{inputenc}
\usepackage{microtype}
\usepackage{inconsolata}

\usepackage{amsmath}
\usepackage{amssymb}
\usepackage{amsfonts}
\usepackage{bbold}
\usepackage{multicol}
\usepackage{multirow} 
\usepackage{pifont}
\usepackage{tabularx}
\usepackage{lipsum} 
\usepackage{longtable}
\usepackage{tabularx} 
\usepackage{array} 
\usepackage{xltabular}
\usepackage{listings}

\usepackage[inline]{enumitem}

\usepackage{booktabs,caption}
\usepackage[flushleft]{threeparttable}
\usepackage[most]{tcolorbox}
\usepackage{makecell}
\tcbuselibrary{breakable}

\newcommand{\code}[1]{{\ttfamily#1}}

\definecolor{mediumgray}{gray}{0.6} 

\DeclareMathAlphabet{\mathpzc}{OT1}{pzc}{m}{it}

\title{\textsc{Mair}: A Massive Benchmark for Evaluating Instructed Retrieval}

\author{Weiwei Sun\textsuperscript{\rm 1} \quad Zhengliang Shi\textsuperscript{\rm 2} \quad Jiulong Wu\textsuperscript{\rm 3} \quad Lingyong Yan\textsuperscript{\rm 4}\\
\textbf{Xinyu Ma}\textsuperscript{\rm 4} \quad \textbf{Yiding Liu}\textsuperscript{\rm 4} \quad \textbf{Min Cao}\textsuperscript{\rm 3} \quad \textbf{Dawei Yin}\textsuperscript{\rm 4}\thanks{~~Corresponding authors} \quad \textbf{Zhaochun Ren}\textsuperscript{\rm 5}\footnotemark[1] 
\\
\textsuperscript{\rm 1}Carnegie Mellon University\quad \textsuperscript{\rm 2}Shandong University \\ \textsuperscript{\rm 3}Soochow University \quad \textsuperscript{\rm 4}Baidu Inc. \quad  \textsuperscript{\rm 5}Leiden University\\
\texttt{\{sunnweiwei,zhengliang.shii,lingyongy,xinyuma2016\}@gmail.com}\\ 
\texttt{yindawei@acm.org,\quad z.ren@liacs.leidenuniv.nl}
}

\begin{document}
\maketitle
\begin{abstract}

Recent information retrieval (IR) models are pre-trained and instruction-tuned on massive datasets and tasks, enabling them to perform well on a wide range of tasks and potentially generalize to unseen tasks with instructions. However, existing IR benchmarks focus on a limited scope of tasks, making them insufficient for evaluating the latest IR models.
In this paper, we propose \textsc{Mair} (Massive Instructed Retrieval Benchmark), a heterogeneous IR benchmark that includes 126 distinct IR tasks across 6 domains, collected from existing datasets. We benchmark state-of-the-art instruction-tuned text embedding models and re-ranking models.
Our experiments reveal that instruction-tuned models generally achieve superior performance compared to non-instruction-tuned models on \textsc{Mair}. Additionally, our results suggest that current instruction-tuned text embedding models and re-ranking models still lack effectiveness in specific long-tail tasks. \textsc{Mair} is publicly available at \url{https://github.com/sunnweiwei/Mair}.

\end{abstract}

\section{Introduction}

Large Language Models (LLMs) have demonstrated impressive capabilities in performing a wide range of natural language processing (NLP) tasks by being first pre-trained on large-scale corpora and then instruction-tuned on numerous downstream 
 tasks~\cite{flant5, selfinstruct, Wang2022SuperNaturalInstructionsGV, flanwei}. This advancement has garnered significant attention in other fields, including Information Retrieval (IR)~\citep{Gao2021UnsupervisedCA, Neelakantan2022TextAC, Wang2022TextEB, Izacard2021UnsupervisedDI, Wang2023ImprovingTE, Asai2022TaskawareRW, Su2022OneEA}.

IR techniques aim to retrieve a set of relevant candidates from a large corpus based on semantic relevance or other query-specific criteria~\cite{yates2021pretrained, fan2022pre}. 
These techniques are critical components of many AI applications, from web search~\cite{zhu2023large} to various retrieval-augmented tasks~\cite{gao2023retrieval, shi-etal-2024-generate}. 
However, most traditional IR models, typically trained on a single task, often exhibit poor generalization to other IR tasks or domains~\cite{Thakur2021BEIRAH, zhao2024dense}. 
Inspired by the success of LLMs, recent research explores training models for the general-purpose IR~\cite{Sachan2022ImprovingPR, sun2023chatgpt, oh2024instructir, weller2024followir}. 
These models, instruction-tuned on multiple retrieval tasks, show significant improvements in aligning with the user intent across different IR tasks.

\begin{table*}[htbp]
\centering
\resizebox{\linewidth}{!}{
\setlength\tabcolsep{3pt}
\begin{tabular}{l ccccc}
\toprule
 & \textsc{Mair} &  BEIR & KILT & FollowIR & InstructIR\\
& (this work)  & \citet{Thakur2021BEIRAH} & \citet{Petroni2020KILTAB} & \citet{weller2024followir} & \citet{oh2024instructir}\\
\midrule
Number of tasks & 126 & 18 & 11 & 3 & 1\\
Number of domains & 6 & 4 & 1 & 1 & 1\\
Number of instructions & 805 & 14 & 5 & 104 & 9,906\\
Number of test queries & 10,038 & 47,233 & 50,736 & 104 & 9,906\\
Number of collections & 426 & 18 & 1 & 3 & 1\\
Total number of docs & 4,274,916 & 38,506,129 & 5,903,530 & 98,312 & 16,072\\

\bottomrule
\end{tabular}
}
\caption{Data statistics of \textsc{Mair} and other relevant IR benchmarks.}
\label{table:compare}
\end{table*}

To evaluate the generalization capabilities of newly emerged IR models, several benchmarks such as BEIR~\cite{Thakur2021BEIRAH}, KILT~\cite{Petroni2020KILTAB}, and MTEB~\cite{Muennighoff2022MTEBMT} have been established recently, compiling a variety of IR tasks.
However, as shown in Table~\ref{table:compare}, these benchmarks either (i) contain a relatively small number of tasks or (ii) feature tasks that are too similar, thus providing limited coverage of the broad IR landscape. 
In contrast, instruction-tuned LLMs have been evaluated on hundreds or thousands of diverse NLP tasks~\citep{Wang2022SuperNaturalInstructionsGV, flant5}.

To fill this gap, this paper introduces \textsc{Mair} (Massive Instructed Retrieval Benchmark), a large-scale IR benchmark consisting of 126 diverse retrieval tasks with 805 distinct instructions to evaluate model generalization on unseen tasks. In \textsc{Mair}, we collect tasks from existing IR datasets, such as those found in (i) SIGIR resource track papers, (ii) tasks in existing benchmarks, (iii) publicly accessible shared tasks in TREC (\url{trec.nist.gov}), and (iv) recent LLM benchmarks. 
Tasks in \textsc{Mair} are elaborately selected to cover various information retrieval requirements in practice, including diverse types of queries and document types, as well as specific relevance criteria.
As a result, a total of 126 unique tasks are selected to constitute the benchmark \textsc{Mair}.

Subsequently, we clean and merge the data of these tasks, and then sample them, resulting in a dataset that includes 10,038 queries and 4,274,916 documents.
We perform the sampling mainly due to the data distribution imbalance.
We validate that the sampled benchmark produces results highly correlated with full-scale testing, achieving a balance between evaluation accuracy and cost.

Finally, we manually annotate retrieval instructions based on the queries and corresponding documents for these IR tasks. Following previous work~\cite{weller2024followir}, these instructions specify the types of queries, documents, and relevance criteria. 
We ultimately annotate 805 distinct instructions, representing 805 different query-document-relevance combinations. 
Some tasks include query-level instructions, meaning each query has a different relevance criterion. 
These annotations make \textsc{Mair} particularly challenging and suitable for evaluating instruction-tuned retrievers in terms of their ability to follow instructions in completing unseen tasks.

Based on \textsc{Mair}, we benchmark various different types of retrieval models, including 
\begin{enumerate*}[label=(\roman*)]
\item sparse retriever,
\item  single-task text embedding models~\citep{Ni2021LargeDE,Izacard2021UnsupervisedDI},
\item  non-instruction-tuned multi-task text embedding models~\citep{Li2023TowardsGT,Xiao2023CPackPR,Wang2022TextEB},
\item  instruction-tuned embedding models~\citep{Wang2023ImprovingTE,Lee2024NVEmbedIT}, and
\item re-ranking models~\citep{sun2023chatgpt}.
\end{enumerate*}
We evaluate on both \textit{no instruction} and \textit{with instruction}, and found instruction-tuned embedding models show clear improvement when instructions are added.
\code{GritLM-7B} achieves the best overall score, with an average nDCG@10 of 55.20.
Furthermore, both \code{e5-mistral-7b-instruct} and \code{GritLM-7B} show notable improvement when instructions are added.


\section{Data Construction}

This section illustrates the process of data collection and benchmark construction.
As aforementioned, the benchmark for instruction-tuned IR models should be capable of evaluating IR baselines on a variety of tasks across different domains and using as realistic instruction as possible.
To this end, we construct our benchmark based on the following criteria: 
\textbf{(a) Task and domain variety}: To assess the generalization of different IR baselines on various tasks from different domains, we collect data from a large range of domains and tasks.
Specifically, the data is collected from 126 distinct IR tasks across 6 domains.
Each task is manually filtered to avoid duplication.
\textbf{(b) Instruction diversity}: For each task, we annotate and review a large number of detailed instructions.
These instructions can assist in thoroughly evaluating models' instruction-following capabilities when searching queries.

\subsection{Data Collection}
To build a comprehensive IR benchmark for instruction-following evaluation, we collect data from the following four well-known sources:

\begin{itemize} 
\item \textbf{Existing IR Resources}: The specialized resource tracks in some information retrieval conferences (e.g. SIGIR).
Specifically, we collect released papers from 2021-2024 SIGIR conferences and finally collect 10 tasks across 2 domains.

\item \textbf{Other IR Benchmarks}: We leverage tasks from existing IR benchmarks such as BEIR~\citep{Thakur2021BEIRAH} and KILT~\citep{Petroni2020KILTAB}, as well as domain-specific benchmarks for evaluating Voyage embedding\footnote{\url{https://blog.voyageai.com/}}. These benchmarks have gained attention in the IR community and provide diverse tasks tailored to specific applications. 
Finally, We collect 55 tasks across 6 domains from the IR benchmarks.

\item \textbf{TREC Tracks}: The \textit{Text REtrieval Conference} (TREC) is a long-standing and well-established IR conference series organized by the National Institute of Standards and Technology (NIST). TREC provides unique and realistic use cases along with rigorously annotated data. Finally, We collect 34 tasks and across 5 domains from TREC tracks. 

\item \textbf{LLM Evaluation Datasets}: In addition to the above datasets, we also integrate several public LLM evaluation datasets released in the LLM era.
Including those datasets can help extending our benchmark to diverse potential instruction-following scenarios.
Finally, We collect 27 tasks across 4 domains from the LLM Evaluation Datasets.
\end{itemize}

Finally, we collect 126 tasks.
There is a simple process for us to collect data:
We first review the documentation of various conference tracks, merging tasks with identical corpus and settings to avoid duplication. 
Then, based on task requirements, we download publicly available datasets from official channels.
For tasks with incomplete corpus, we use web crawling and other techniques to supplement the data as much as possible. 
Since the data sources are diverse and the original formats vary substantially, we perform necessary data cleaning operations such as deduplication, keyword extraction, and text normalization on the data. Thus, we unify these task data into a consistent format.






\begin{figure}[!t]
 \centering
\includegraphics[width=1\columnwidth]{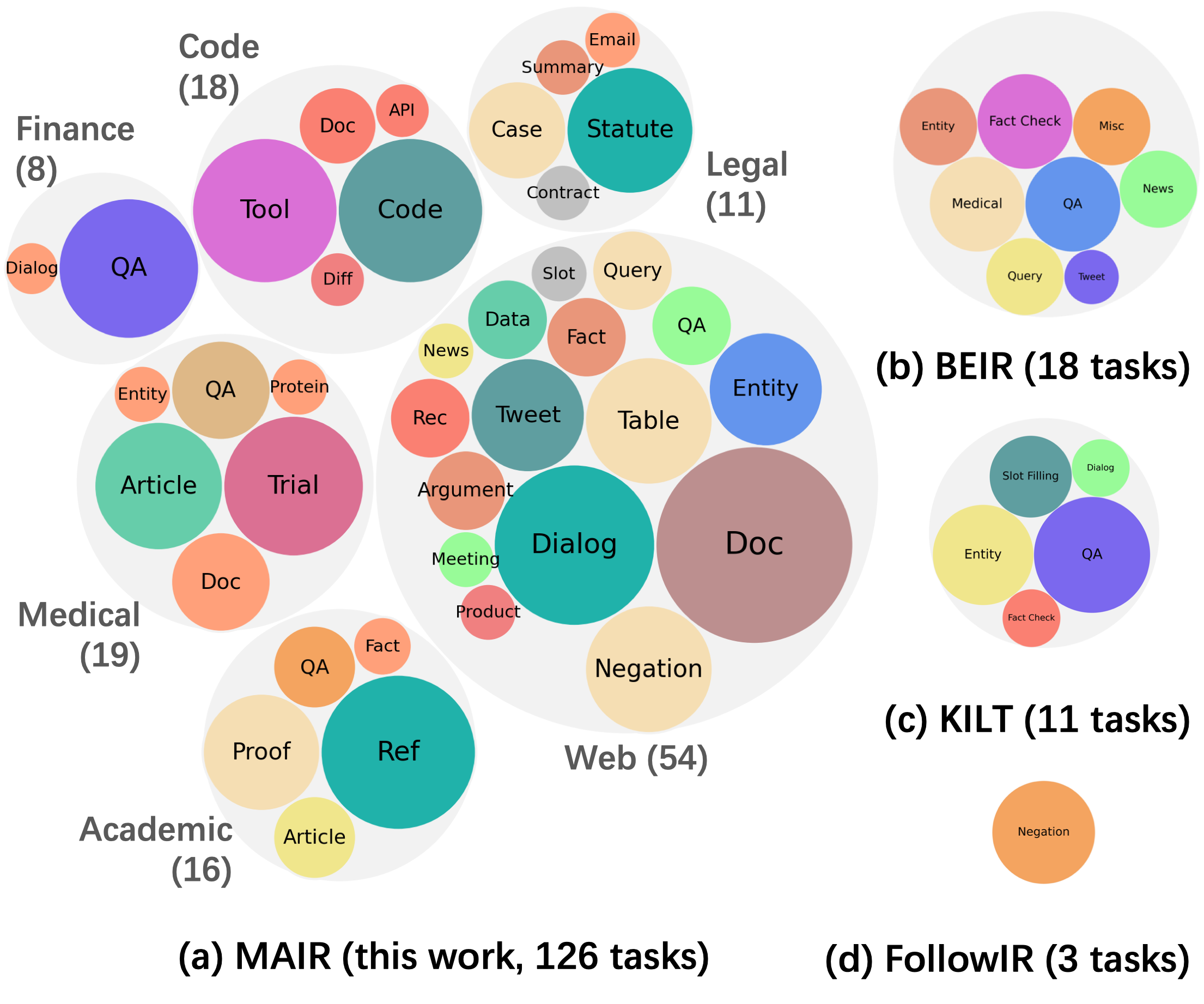} 
\caption{Compared to other datasets, \textsc{Mair} covers a more diverse types of task. Bubble size represents the number of tasks of each type.}
\label{fig:bubble}
\end{figure}



\subsection{Data Sampling}
After filtering out 126 tasks, we find the data distributions are quite imbalanced among different tasks.
Besides, since the scales of some datasets are extremely large with amounts of redundant content,  model evaluation of the whole dataset is inefficient and unnecessary.
To alleviate the influence of task data imbalance and improve evaluation efficiency, we lightweight our benchmark while preserving its evaluative capability via effective data sampling.

Specifically, for each task, we reduce its sample size following the following two steps:
\begin{itemize}
    \item \textbf{Query Sampling}: First, for all search queries in the task, we perform the K-means algorithm over query embeddings to cluster the queries. And we randomly sample one single query in each cluster to ensure query diversity while avoiding redundancy. And we set the cluster number as 100 for each task. In this way, we can finally sample 100 queries for each task.
    \item \textbf{Document Sampling}: Second, for each sampled query above, we use all originally annotated documents for evaluation (for example, each document is labeled related or not). Since there are some queries will few annotated documents, we randomly sample some unlabeled documents as negative documents for these queries. 
\end{itemize}

It is noteworthy that some tasks may be originally evaluated on too small size of corpus making it difficult to sample enough documents in the above step 2. To address this issue, we manually combine the documents together if a small task is similar to a larger task. As a result, both tasks can share the same large-scale corpus.

The resulting benchmark comprises 10,038 queries and 4,274,916 documents (about 2 billion tokens based on OpenAI cl32k tokenizer), providing a computationally efficient yet representative testbed for IR models. 

\subsection{Instruction Annotation}
After collecting the tasks above, we manually write the retrieval instructions for each task.
All the instructions are written and reviewed by the experts in information retrieval.
Following \citet{Asai2022TaskawareRW}, the basic format of the instruction describes the query, the passages, and the relevance criterion. For example, instruction of \code{ClinicalTrials} is 
``\textit{Given a patient descriptions, retrieve clinical trials that suitable for that patient. The patient description (query) is a questionnaire that is filled by the patient or their clinician. Trials are relevant if the patient met the
inclusion criteria and did not meet any exclusion criteria; Trials are partially relevant if the patient met the
inclusion criteria but was excluded by one or more exclusion criteria.}'',.

For tasks where different search queries require unique instructions, we provide query-level annotations.
The following are some examples of instructions besides the basic format:
For example, task \code{Genomics-AdHoc\_2007}~\citep{Hersh2007TREC2G} aims to retrieve passages that contain a specific type of biomedical entity eg. antibodies, proteins, and strains. We annotate the entity type and its definition in instructions for each query.


After the above steps, we obtain the final datasets, which consist of 10,038 queries from 126 IR tasks, with 805 instructions annotated for each task. There are in total 426 document collections and 4,274,916 documents across various domains, such as news articles, scientific papers, web pages, and code repositories. These datasets contain well-designed instructions and various document collections. They can serve as a comprehensive benchmark to evaluate the ability of retrieval systems in understanding natural language instructions and retrieving relevant information from different sources.
See

\section{Dataset Analysis}

The Table~\ref{table:list-tasks-1} and \ref{table:list-tasks-2} lists the full list of the tasks in \textsc{Mair} and their input / output, and task type and domain. Tasks in \textsc{Mair} mainly come from six domains: (i) Web refers to retrieving information from the general web. 47  IR tasks are included in the web domain.
(ii) Academic domain focus on retrieving academic literature or retrieval for academic applications. (iii) The code domain consists of 18 tasks, ranging from code retrieval to tool retrieval, and to code agent. (iv) Legal domain focuses on legal-related tasks, such as case retrieval and statute retrieval. (v) Finance domain consists of 8 tasks that concern IR application in Finance task. (vi) Medical domain mainly consists of the shared tasks in \textsc{TREC-CDS} and \textsc{Genomics}, which wide range of retrieval and matching tasks in biomedical.

Figure~\ref{fig:bubble} highlights the domain and task categories of \textsc{Mair}, and compares them with previous datasets. 
\textsc{Mair} covers more diverse domains and types of retrieval tasks.
Next, we analyze the correlation between different tasks to further study the task diversity of \textsc{Mair}, and validate the effectiveness of the data sampling approaches.

\begin{figure}[!t]
 \centering
\includegraphics[width=1\columnwidth]{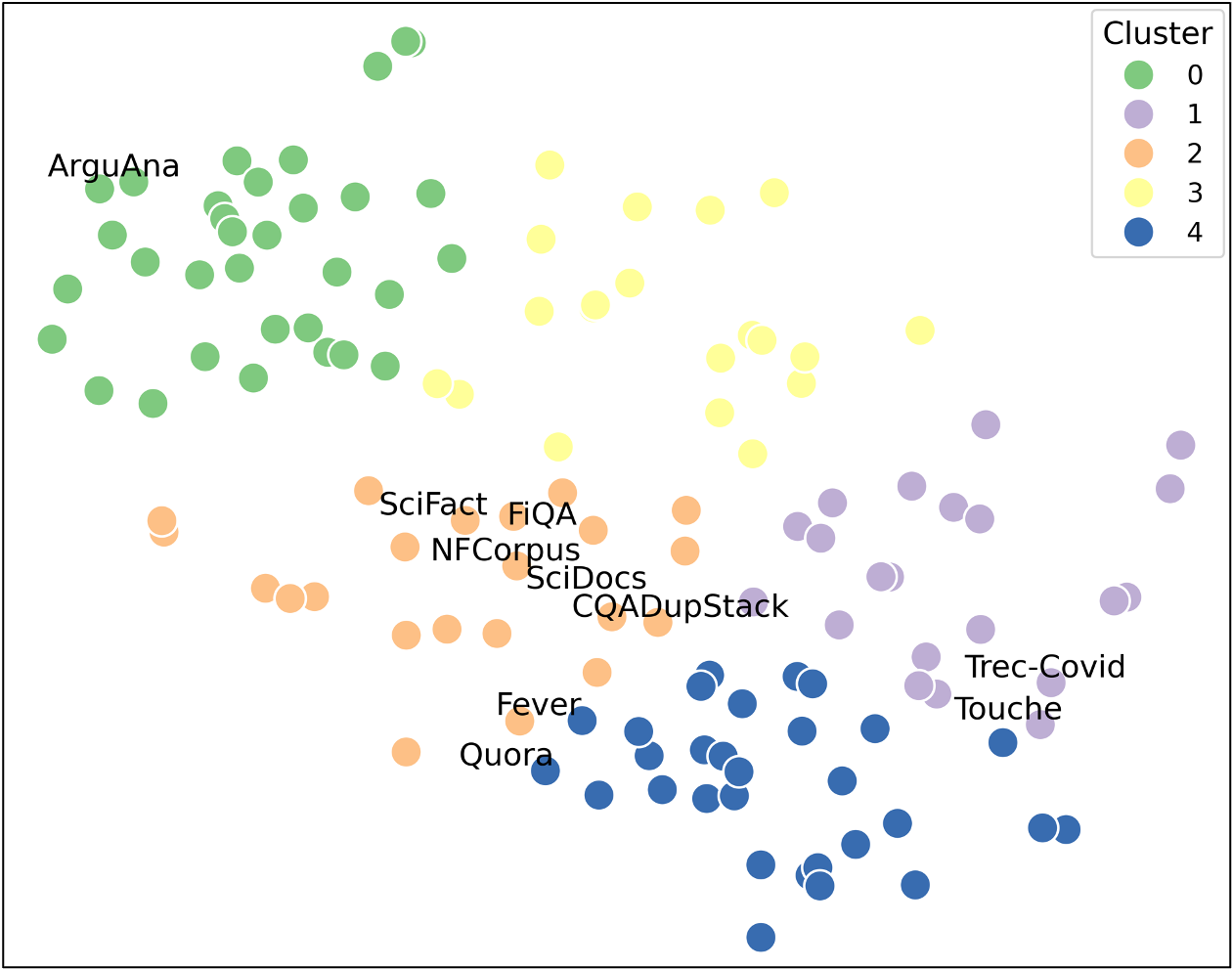} 
\caption{Visualization of the correlation among 126 tasks in \textsc{Mair}, with annotations for tasks from BEIR. \textsc{Mair} includes more diverse tasks. Task similarity is determined based on the performance correlation of all baseline models. We employ KMeans for clustering and t-SNE for visualization.}
\label{fig:cluster}
\end{figure}

\paragraph{Task Correlation}
To measure the task diversity of \textsc{Mair}, we 
calculate the similarity of two tasks using the Pearson correlation coefficient of the different models' performance on two tasks. 
Based on the results of all baseline models on the \textsc{Mair}, we calculated the correlation between all tasks, and got a correlation matrix in $\mathbf{M}=\mathbb{R}^{126\times126}$, $\mathbf{M}[i,j]$ denote correlation between task $i$ and $j$. We plot this matrix in Figure~\ref{fig:heat}. 
To better visualize the matrix, we use t-SNE to visualize the task correlation matrix in 2D.

Figure~\ref{fig:cluster} presents the t-SNE visualization of the correlation matrix, where each task is represented as a point, and the distance between two tasks reflects their correlation.
We have annotated the tasks from BEIR and observed that most of these tasks cluster together in the orange group, indicating that the performance of different models on these tasks is highly correlated. 
In contrast, tasks in \textsc{Mair} show greater diversity, covering a larger area and thus providing a more comprehensive evaluation of the models. 
The specific pairwise correlations are displayed in the heatmap in Figure~\ref{fig:heat}, revealing that many tasks exhibit negative correlations. This suggests that these tasks have significantly different definitions, leading models that perform well on one task to perform poorly on another.

\begin{figure}[!t]
 \centering
\includegraphics[width=1\columnwidth]{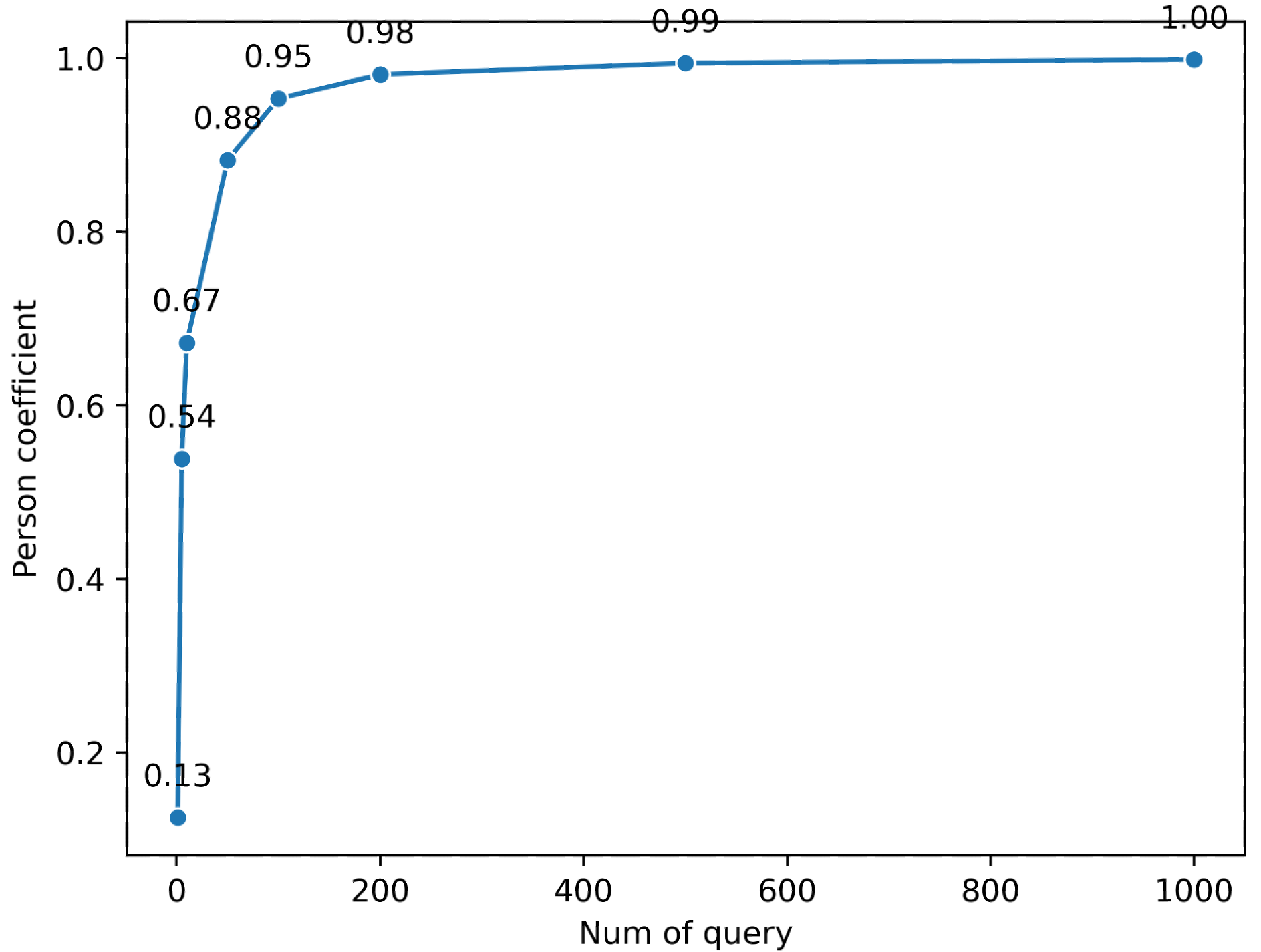} 
\caption{The performance correlation of baseline models with different sampled numbers of queries. Sampling 100 queries achieves a good trade-off between correlation and cost.}
\label{fig:line}
\vspace{-1em}
\end{figure}

\paragraph{Sampling Effectiveness}
In our data construction process, we sample the data to obtain a lightweight test set. However, the sampled data may lead to bias in the evaluation results.
To measure the effectiveness of our data sampling process, we build a test set with different maximum test sizes cut off, ranging from [1, 5, 10, 50, 100, 200, 500, 10000]. Then, we run the baseline models on these test sets, can compute the task correlation between the full test set (i.e., unsampled test set), and the sampled test set.
A high correlation means that evaluation results on the sampled test set are very similar to the evaluation on the full set.
Figure~\ref{fig:line} illustrates the correlation of different test sets cut off.
We can see that retaining more instances leads to a higher correlation, i.e., the evaluation is more robust.
Notably, a cut-off of $100$ instances achieves over $0.95$ Pearson correlation coefficient between the evaluation results, indicating that our sampling method could achieve good reliability while using minimal cost.

\begin{table*}[ht]
\centering\small
\setlength\tabcolsep{2pt}
\begin{tabular}{l cc cc cc cc cc cc cc}

\toprule
\textbf{Model} & 
\multicolumn{2}{c}{\textbf{Avg}} & 
\multicolumn{2}{c}{\textbf{Web}} & 
\multicolumn{2}{c}{\textbf{Academic}} & 
\multicolumn{2}{c}{\textbf{Legal}} & 
\multicolumn{2}{c}{\textbf{Medical}} & 
\multicolumn{2}{c}{\textbf{Finance}} & 
\multicolumn{2}{c}{\textbf{Code}}\\
\midrule
Instruction?  & \ding{55} & \ding{51} & \ding{55} & \ding{51}& \ding{55} & \ding{51}& \ding{55} & \ding{51}& \ding{55} & \ding{51}& \ding{55} & \ding{51}& \ding{55} & \ding{51} \\
\midrule

BM25 & 40.75 & - & 40.21 & - & 38.50 & - & 47.60 & - & 46.01 & - & 51.22 & - & 32.72 & -   \\
\midrule
\code{contriever-msmarco} & 39.88 & - & 45.19 & - & 31.82 & - & 31.74 & - & 36.60 & - & 50.59 & - & 35.06 & -   \\

\code{gtr-t5-base} & 37.79 & - & 40.55 & - & 31.36 & - & 33.77 & - & 32.95 & - & 47.78 & - & 37.18 & -  \\

\code{gtr-t5-large} & 41.43 & - & 43.97 & - & 35.62 & - & 37.35 & - & 33.93 & - & 51.83 & - & 42.29 & -  \\

\midrule

\code{all-MiniLM-L6-v2} & 38.66 & 26.90 & 40.05 & 25.56 & 42.90 & 35.85 & 29.20 & 18.46 & 35.24 & 22.32 & 44.54 & 28.15 & 35.90 & 29.14   \\

\code{e5-small-v2} & 41.36 & 34.47 & 43.24 & 34.12 & 38.83 & 32.83 & 32.20 & 28.55 & 37.89 & 30.82 & 55.32 & 50.49 & 39.70 & 35.26    \\

\code{e5-base-v2} & 43.45 & 38.80 & 43.78 & 38.95 & 42.00 & 38.37 & 35.76 & 33.06 & 40.22 & 30.68 & 58.33 & 54.18 & 43.48 & 40.13   \\

\code{e5-large-v2} & 44.75 & 39.45 & 44.60 & 38.93 & 45.44 & 40.01 & 36.96 & 32.86 & 41.90 & 35.13 & 58.92 & 54.27 & 44.28 & 40.00  \\

\code{gte-base-en-v1.5} & 44.06 & 34.25 & 44.72 & 30.97 & 40.15 & 35.34 & 37.89 & 34.73 & 45.63 & 37.18 & 61.64 & 49.23 & 40.50 & 33.11   \\

\code{gte-large-en-v1.5} & 46.58 & 33.87 & 46.38 & 30.97 & 47.00 & 40.23 & 41.04 & 34.25 & 47.75 & 32.91 & 63.47 & 41.57 & 41.58 & 32.66  \\

\code{bge-base-en-v1.5} & 42.58 & 30.63 & 43.33 & 27.70 & 38.85 & 31.66 & 33.56 & 26.89 & 43.78 & 31.58 & 55.54 & 38.95 & 42.46 & 35.05  \\

\code{bge-base-en-v1.5} & 44.34 & 28.01 & 44.91 & 24.78 & 41.20 & 32.05 & 37.35 & 24.09 & 45.31 & 25.71 & 58.69 & 31.38 & 42.58 & 34.77  \\

\code{text-embedding-3-small} & 47.89 & 45.28 & 48.10 & 46.65 & 44.43 & 41.11 & 45.71 & 41.06 & 42.74 & 37.81 & 61.36 & 57.99 & 48.87 & 46.74  \\

\midrule

\code{gte-Qwen2-1.5B-instruct}$^\spadesuit$ & 49.14 & 51.81 & 46.81 & 51.30 & 48.87 & 49.98 & 50.91 & 51.34 & 45.97 & 45.20 & 62.85 & 64.61 & 50.40 & 53.48 \\

\code{NV-Embed-v1}$^\spadesuit$ & 50.29 & 51.19 & 49.76 & 51.07 & 50.70 & 51.14 & 44.07 & 44.57 & 40.71 & 38.96 & 68.27 & 69.59 & 52.62 & 54.52  \\

\code{e5-mistral-7b-instruct}$^\spadesuit$ & 50.85 & 54.43 & 48.52 & \textbf{54.57} & 48.78 & 50.03 & \textbf{52.94} & 52.50 & \textbf{48.35} & \textbf{51.10} & 66.52 & 68.76 & 52.29 & 54.83 \\

\code{GritLM-7B}$^\spadesuit$ & 48.46 & \textbf{55.26} & 43.01 & 54.50 & 51.46 & \textbf{53.45} & 50.70 & \textbf{53.14} & 45.38 & 48.22 & 64.28 & \textbf{70.83} & 53.62 & \textbf{57.52}  \\

\midrule
\code{monot5-base-msmarco} & 43.51 & 38.27 & 47.03 & 42.24 & 35.44 & 27.00 & 41.45 & 38.41 & 36.19 & 31.31 & 59.34 & 56.72 & 40.27 & 34.25  \\

\code{mxbai-rerank-large-v1} & 42.84 & 22.91 & 44.33 & 24.94 & 32.98 & 15.53 & 42.50 & 19.14 & 43.25 & 24.64 & 56.54 & 26.23 & 41.78 & 23.78 \\

\code{jina-reranker-v2-base} & 50.88 & 48.59 & \textbf{52.31} & 50.74 & 45.22 & 42.96 & 49.32 & 46.10 & 46.78 & 42.87 & 61.35 & 59.28 & 51.03 & 48.24   \\

\code{bge-reranker-v2-m3} & 46.59 & 40.45 & 48.63 & 43.64 & 40.01 & 33.38 & 49.18 & 42.38 & 41.12 & 30.85 & 64.17 & 60.36 & 41.47 & 34.67  \\

\code{bge-reranker-v2-gemma} & \textbf{52.20} & 38.16 & 51.26 & 42.40 & \textbf{52.09} & 30.89 & 47.74 & 30.47 & 45.69 & 28.96 & \textbf{68.34} & 46.41 & \textbf{54.00} & 39.86  \\

\code{gpt-3.5-turbo}$^\spadesuit$  & 47.84 & 49.02 & 47.80 & 49.70 & 41.59 & 41.02 & 47.47 & 48.38 & 42.21 & 41.97 & 64.45 & 64.52 & 49.93 & 52.27  \\
\bottomrule
\end{tabular}
\caption{\textbf{Main Results (nDCG@10)} Rows marked with \ding{51} indicate results with instruction input, while rows marked with \ding{55} indicate results without instruction input. Models labeled with $^\spadesuit$ are instruction fine-tuned.  The last group represents the re-ranking models, all of which re-rank the top 100 results from \code{text-embedding-3-small}.}
\label{table:main-results}
\end{table*}

\section{Experimental Setup}

\subsection{Models}

\paragraph{Sparse Retrieval} These measure relevance by computing term overlap. We benchmark BM25, implemented in BM25S~\citep{L2024BM25SOO}.

\paragraph{Single-Task Text Embedding} These models are trained on a single IR dataset. We benchmark \code{gtr-t5-base}, \code{gtr-t5-large}, and \code{contriever-msmarco}, all of which are trained on MS MARCO~\citep{Ni2021LargeDE,Izacard2021UnsupervisedDI}.

\paragraph{Non-Instruction-Tuned Multi-Task Text Embedding} These models are typically trained on various annotated IR training datasets combined with massive weekly supervised data. We benchmark (i) the GTE series~\citep{Li2023TowardsGT}: \code{gte-base-en-v1.5}, \code{gte-large-en-v1.5}; (ii) the BGE series~\citep{Xiao2023CPackPR}: \code{bge-base-en-v1.5}, \code{bge-large-en-v1.5}; (iii) the E5 series~\citep{Wang2022TextEB}: \code{e5-small-v2}, \code{e5-base-v2}, \code{e5-large-v2}; and (iv) \code{all-MiniLM-L6-v2} from Sentence Transformer. We also evaluated \code{text-embedding-3-small} by the OpenAI API\footnote{\url{https://platform.openai.com/docs/guides/embeddings}}.

\paragraph{Instruction-Tuned Text Embedding} 
These models are fine-tuned on instruction datasets, where the model input is a query paired with an instruction describing the retrieval task. We benchmark (i) \code{e5-mistral-instruct}, which optimizes instruction-following ability using LLM-generated data~\citep{Wang2023ImprovingTE}; (ii) \code{NV-Embed-v1}, which utilizes bidirectional attention with an additional latent attention layer to enhance text embedding model~\citep{Lee2024NVEmbedIT}; (iii) \code{GritLM-7B}~\citep{Muennighoff2024GenerativeRI}, a unified text embedding and generation model; and (iv) \code{gte-Qwen2-1.5B-instruct}~\citep{Li2023TowardsGT}, general text embedding based on Qwen2-1.5B.

\paragraph{Cross-Encoder Re-Rankers} These models measure the relevance of paired queries and passages using bidirectional or unidirectional Transformers. We benchmark (i) \code{monoT5-Base}, a T5 encoder model trained on MS MARCO,  (ii) \code{mxbai-rerank-large-v1} and \code{jina-reranker-v2-base}, multi-task reranker developed by Mxbai and Jina.ai, respectively, and (iii) \code{bge-reranker-v2-m3} and \code{bge-reranker-v2-gemma}, trained on massive ranking data with XLM-R and Gemma-2B as their respective backbones.

\paragraph{LLM-based Re-Rankers} These models prompt general-purpose LLMs to perform re-ranking in a zero-shot setting. We benchmark RankGPT with the \code{gpt-3.5-turbo}~\citep{sun2023chatgpt}.

\begin{figure}[!t]
 \centering
\includegraphics[width=1\columnwidth]{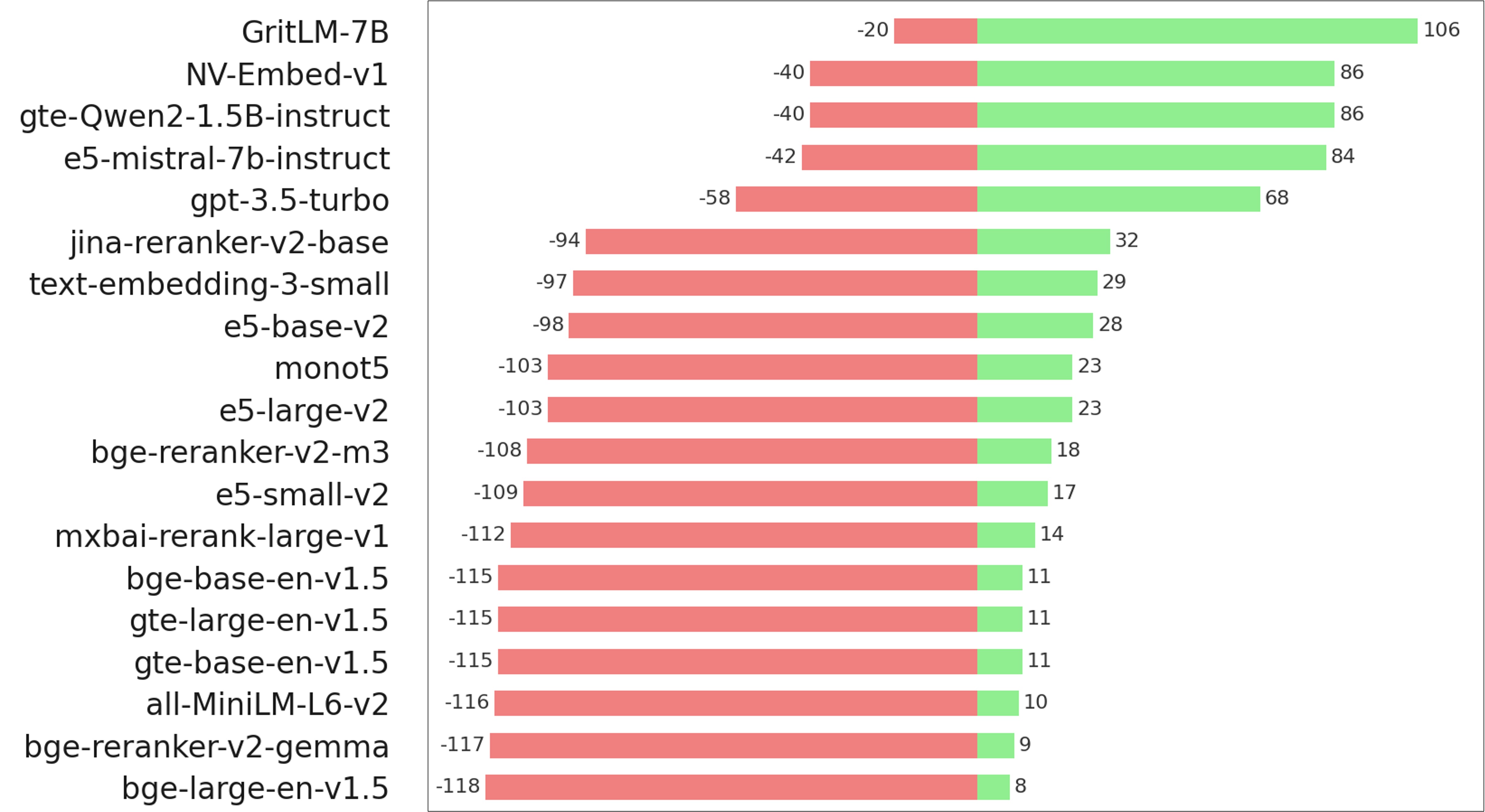} 
\caption{With the addition of instruction, the number of tasks that obtain performance improvement (green part) and reduction (red part). We can see that instruction-tuned models show more improvements while non-instruction-tuned models reduce on most tasks.}
\label{fig:bar}
\vspace{-1em}
\end{figure}

\subsection{Evaluation}
Following previous work, we use \textbf{nDCG@10} as the evaluation metric. 
The overall score is defined as the average score across all queries. We also report the average nDCG@10 for each of the following domains: Web, Academic, Code, Medical, Legal, and Finance. The specific tasks included in each field can be found in Table \ref{table:all-results-1} - \ref{table:all-results-5}. 

For all models, we consider two settings: (i) \textit{no instruction}, which retrieves passages without task instruction or with a simple web search instruction when required, and (ii) \textit{+ instruction}, which retrieves passages with instruction input paired with the query. The performance changes after using instructions indicate the model's ability to understand them.
Note that for non-instruction-tuned models, we also test their performance under the instruction setting for reference, even though they may not be optimized to understand instructions.

All re-ranking models use \code{text-embedding-3-small} as the first-stage retriever and re-rank the top-100 passages. Passages are truncated to the maximum input length of each model.

\begin{figure*}[!t]
 \centering
\includegraphics[width=1.75\columnwidth]{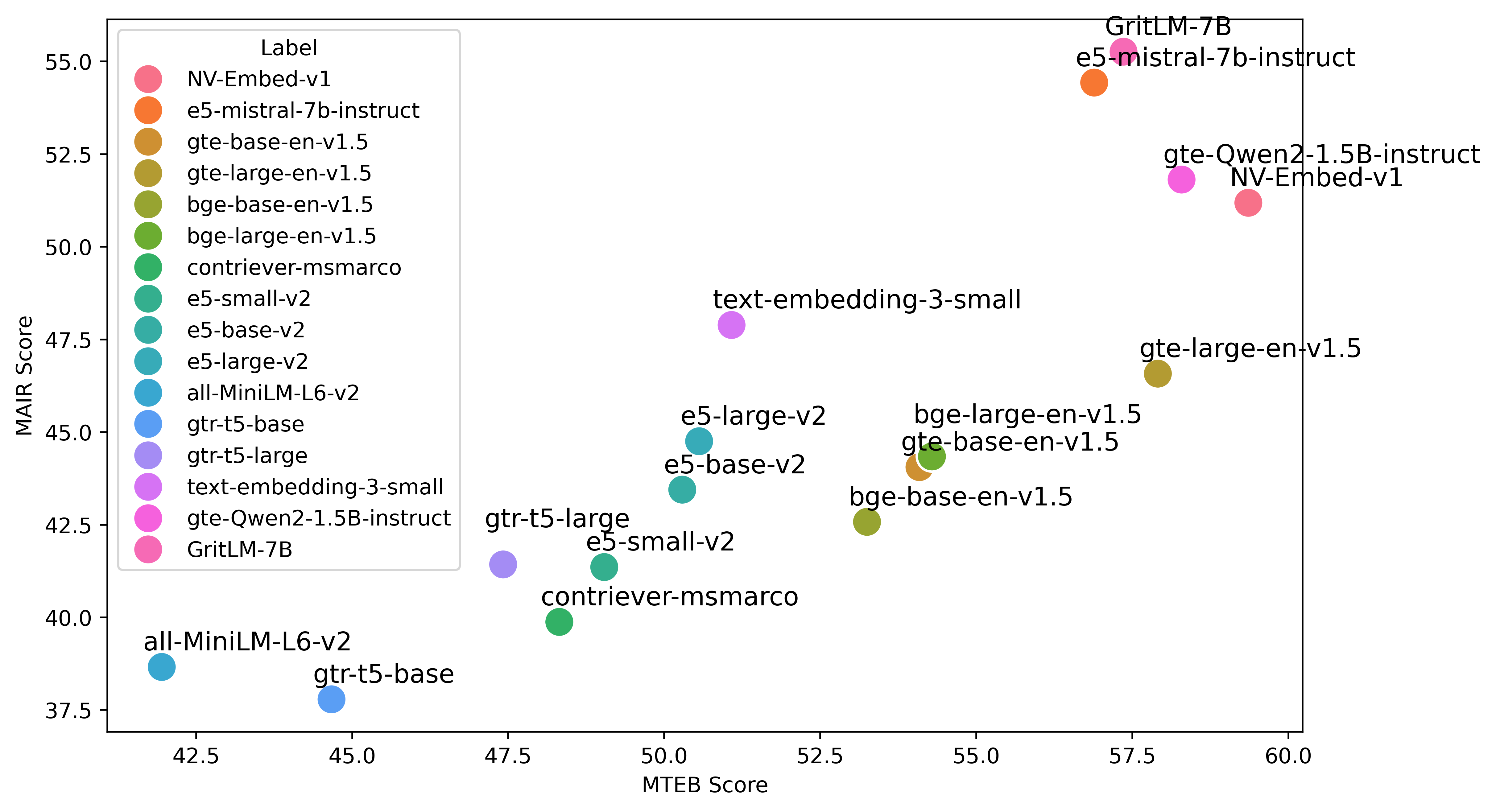} 
\caption{Score between MTEB (Retrieval) and \textsc{Mair}.}
\label{fig:mteb}
\end{figure*}

\section{Evaluation Results}

\subsection{Main Results}

Table~\ref{table:main-results} reports the evaluation results of the tested models. Table~\ref{table:all-results-1}-\ref{table:all-results-5} reports the detailed results of each individual task.
We observe that instruction-tuned embedding models with instruction input achieve the best overall performance. \code{GritLM-7B} achieves an average score of 48.40, with a 6.80 nDCG improvement when instructions are added. \code{NV-Embed} also shows a clear improvement with the addition of instructions, though not as significant as \code{GritLM-7B}. This difference is likely due to the LLM-generated instruction-tuning data enhancing the models' ability to understand instructions~\citep{Wang2023ImprovingTE}.

Non-instruction-tuned models experience a performance drop when instructions are added. Some, like \code{text-embedding-3-small}, show a slight decrease, while others, such as \code{bge-base-en-v1.5}, exhibit a more significant decline.

For the re-ranker, \code{bge-reranker-v2-gemma} achieves the best results, outperforming all embedding models when no instructions are provided, but it shows a notable decline in performance when instructions are added. 
RankGPT based on \code{gpt-3.5-turbo} achieves results close to \code{bge-reranker-v2-m3} without instruction input but demonstrates a 1.21 nDCG improvement when instructions are included. 
This suggests that prompted LLMs can intuitively transfer general instruction-following capabilities to ranking tasks.

\subsection{Gain of Instruction}

Figure~\ref{fig:bar} shows that with the addition of instructions, the model’s performance improves on a number of the 126 tasks while decreasing on others. We observe that the instruction-tuned models show improvement on more than half of the tasks, with \code{GritLM-7B} performing the best, achieving improvements on 106 out of 126 tasks. \code{NV-Embed-v1}, \code{gte-Qwen2-1.5B-instruct}, and \code{e5-mistral-7b-instruct} achieve similar results and outperform \code{gpt-3.5-turbo}. In contrast, non-instruction-tuned models show a performance decrease on more tasks when instructions are added.

\subsection{Compare with MTEB}
Figure~\ref{fig:mteb} shows the relationship between models' performance on MTEB (Retrieval) and \textsc{Mair}. We can see that the two benchmarks share a similar trend. Among these models, \code{text-embedding-3-small} achieves better results on \textsc{Mair} than on MTEB. 
For example, on MTEB, \code{gte-large-en-v1.5} outperforms \code{text-embedding-3-small} by about 7 points and outperforms \code{e5-mistral-7b-instuct} by 1 point. However, it performs worse than them on \textsc{Mair}. 
This is probably because \code{text-embedding-3-small} has better generalization, as it performs better on massive unseen tasks in \textsc{Mair}, while \code{gte-large-en-v1.5} is more optimized towards tasks in MTEB. 
We also observe that single-task models such as \code{contriever-msmarco} perform poorly on \textsc{Mair}, which indicates that \textsc{Mair} requires more generalization ability.

\begin{figure}[!t]
 \centering
\includegraphics[width=1\columnwidth]{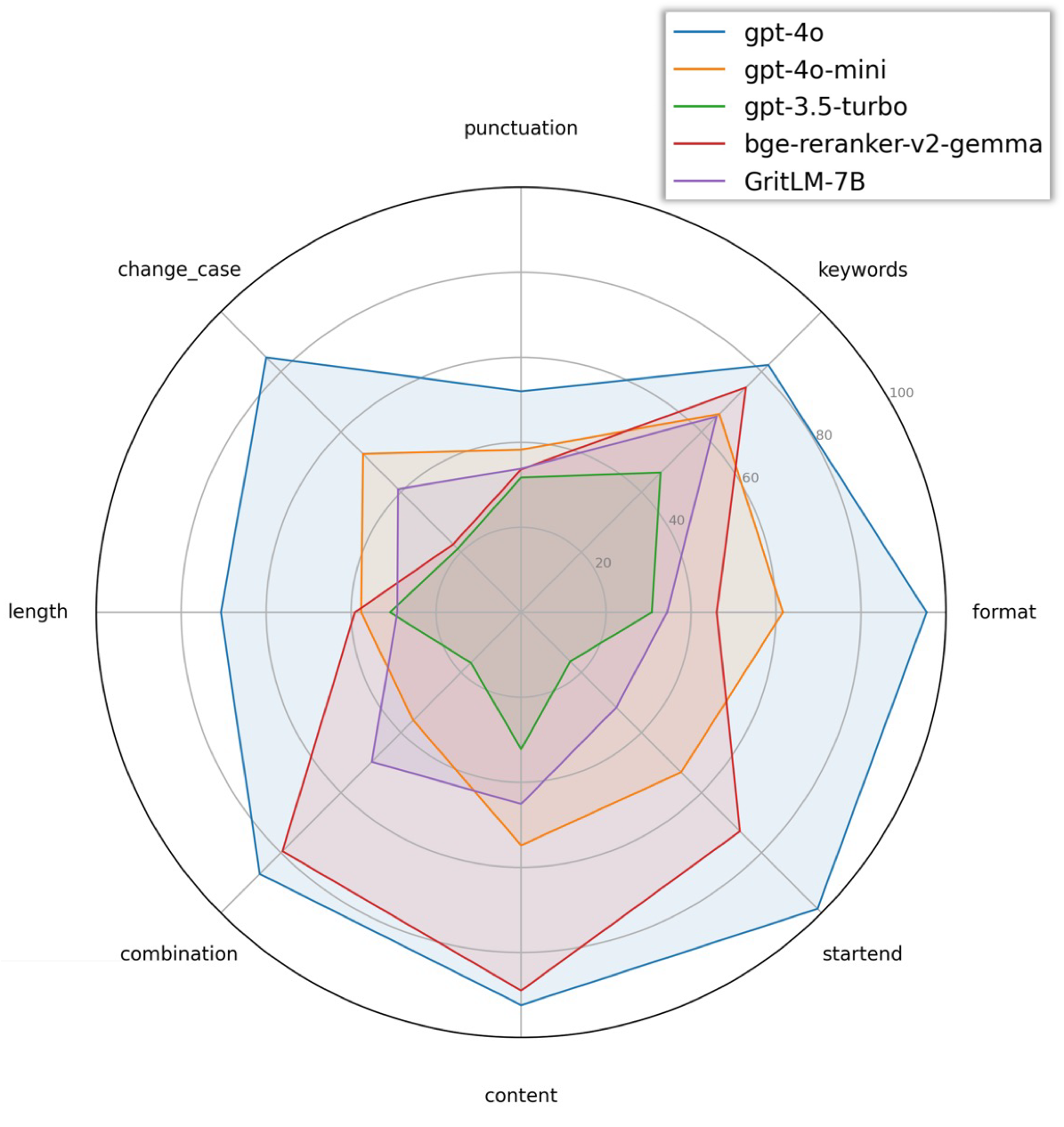} 
\caption{Results (nDCG@10) on IFEval sub-tasks.}
\label{fig:ifeval}
\end{figure}

\subsection{Analysis on IFEval}
To evaluate the model on challenging instruction-following tasks, we designed the IFEval task~\citep{Zhou2023InstructionFollowingEF} within \textsc{Mair}. 
IFEval consists of 8 different instruction-following subtasks, such as \code{format} (selecting responses in a specific format), \code{keywords} (including specific words), and \code{length} (adhering to length restrictions).
The retrieval task in IFEval is to select the answer that correctly follows the instructions from among the 100 candidates.
Specifically, building on the original IFEval data~\citep{Zhou2023InstructionFollowingEF}, we used \code{gpt-3.5-turbo} to generate 100 candidate answers for each question, ensuring that only 10 fully follow the given instructions.
IFEval is challenging for retrieval models because (i) the instructions are out-of-training-distribution; (ii) all candidate answers are semantically relevant to the question, thus, the model must focus on the instructions to identify the correct answer.

Figure~\ref{fig:ifeval} demonstrates the performance of \code{GritLM-7B}, \code{bge-reranker-v2-gemma}, and RankGPT with \code{gpt-3.5-turbo}, \code{gpt-4o-mini}, and \code{gpt-4o} on the 8 subtasks of IFEval. 
The results show that existing instruction-tuned retrieval models still perform poorly on challenge instruction-followed ranking task.
For example, the \code{GritLM-7B} achieves an nDCG@10 score of less than 60 on 7 out of the 8 tasks. 
In contrast, advanced LLM \code{gpt-4o} achieves an nDCG@10 score of over 80 on 6 out of the 8 tasks.
It indicates again the shortcomings of instruction-tuned retrieval models in handling complex information requirements, and utilizing advanced language models as supervisors might be an effective strategy.
For full results on IFEval, refer to Tables~\ref{table:if-eval} and \ref{table:all-sub-results-1}.

\section{Related Work and Background}

\subsection{Instruction tuning for Retrieval}
Instruction tuning has been a effective technique in the development of LLMs~\cite{flant5,selfinstruct,Wang2022SuperNaturalInstructionsGV}. 
In this process, models are trained on diverse instruction-response pairs, empowering them with the ability to adaptively perform unseen tasks based on instructions~\citep{chen2024alpagasus,flanwei}.
This has inspired emergent research on training instruction-tuned retrieval models. 
\citet{Asai2022TaskawareRW,Su2022OneEA} are the earliest works in this direction, where they propose training text embedding models with instructions alongside the query, enabling the models to perform well on unseen tasks using instructions. 
Recent research has started to finetune larger text embedding models with instructions, such as Mistral-7B and Mixtral-7x8B~\citep{Muennighoff2024GenerativeRI}, while \citet{Wang2023ImprovingTE} further proposes using LLMs to generate instruction-tuning data for training retrievers. These instruction-tuned embeddings have climbed to the top of existing IR leaderboards like MTEB~\citep{Muennighoff2022MTEBMT}.
Meanwhile, some paper explore prompt-based method to instruct general-purpose LLMs for re-ranking tasks~\citep{sun2023chatgpt}.

\subsection{IR Benchmark}
Benchmarking has been a crucial part of IR system development. Traditional IR benchmarks typically evaluate models on narrow tasks, like question-answering, or on a certain domain~\cite{Campos2016MSMA, Karpukhin2020DensePR}. Recently, with the advance of pre-trained language models and language model-based retrievers, such as monoBERT~\citep{Nogueira2019PassageRW} and GTR~\citep{Ni2021LargeDE}, increasing attention has been paid to constructing multi-task IR benchmarks and evaluating retrievers on out-of-domain tasks.
A typical effort is BEIR~\citep{Thakur2021BEIRAH}, which consists of 18 tasks across 4 domains. MTEB~\citep{Muennighoff2022MTEBMT} further extends BEIR by adding other embedding-related non-retrieval tasks like clustering and classification. KILT~\citep{Petroni2020KILTAB} collects 11 knowledge-intensive language tasks.
However, given the recent progress of instruction-tuned retrievers, existing benchmarks cannot comprehensively evaluate models' abilities since the number of tasks and the scope of tasks are limited. Some recent papers, such as FollowIR~\citep{weller2024followir} and InstructIR~\citep{oh2024instructir}, propose benchmarks that specifically focus on evaluating models' instruction-following abilities. FollowIR rewrites the original narratives in three TREC shared tasks to include some exclusionary instructions, making some relevant passages become irrelevant. InstructIR utilizes LLM to generate synthetic user backgrounds and further rewrites the relevant passages using LLM to fit the background.
However, these tasks are synthetic and only include limited tasks, making it difficult to robustly evaluate the performance of models on real-world unseen instructions. In comparison, \textsc{Mair} is a large-scale multi-task IR benchmark consisting of 126 distinct tasks. Most tasks are long-tail and without training data, and detailed instructions for each task are manually annotated.

\section{Conclusion}
In this paper, we introduce a novel massive instructed IR benchmark called \textsc{Mair} for evaluating instruction-tuned retrieval models.
Compared with existing related benchmarks, \textsc{Mair} has a more comprehensive coverage of various IR tasks, with different types of queries, documents and relevance criteria. Specifically, \textsc{Mair} comprises 126 distinct IR tasks, with 805 newly annotated instructions.
All tasks and instructions are manually selected and annotated to ensure the benchmark can assess the instruction-following capabilities of different IR models in practice.

Based on \textsc{Mair}, we benchmark various retrieval models, including both instruction-tuned and non-instruction-tuned models. Our results demonstrate that \textsc{Mair} poses greater challenges than existing benchmarks, and provides a more comprehensive evaluation.


\section*{Limitation}
The limitations of this work include the lack of study of retrieval in a multilingual setting. Our benchmark focuses only on the English language and considers only text retrieval. Therefore, we plan to explore multilingual IR settings in the future. Another limitation is the lack of study on prompt sensitivity. It is well known that LLMs are sensitive to prompt words. We plan to annotate more instructions in the future to study how LLM performance is impacted by prompt words.

\section*{Ethics Statement}

We acknowledge the importance of the ACM Code of Ethics and fully agree with it. We ensure that this work is compatible with the provided code in terms of publicly accessible datasets and models. Risks and harms associated with large language models include the generation of harmful, offensive, or biased content. The new benchmark is composed of various previous datasets and is therefore licensed under their respective data licenses.

\newpage
\bibliography{anthology,custom,TREC,SIGIR}
\bibliographystyle{acl_natbib}

\onecolumn

\appendix

\section{Data Card}
The code are available at \url{https://github.com/sunnweiwei/MAIR}.
The dataset are available at \url{https://huggingface.co/datasets/MAIR-Bench/MAIR-Queries} and \url{https://huggingface.co/datasets/MAIR-Bench/MAIR-Docs}.
Please refer to \url{https://github.com/sunnweiwei/MAIR/blob/main/MAIR-Example.pdf} for data processing details and examples of \textsc{Mair}.

\begin{table*}[ht]
\centering\small
\setlength\tabcolsep{1.5pt}
\caption{List of tasks in \textsc{Mair}.}
\label{table:list-tasks-1}
\begin{tabular}{l cccccc}
\toprule
\textbf{Task} & Reference & Query & Doc & Task Type & Domain\\
\midrule
Competition-Math &  \citet{Hendrycks2021MeasuringMP} & Math question & Answer & Question Answering & Academic\\
ProofWiki\_Proof & \citet{Welleck2021NaturalProofsMT} & Math Statement & Proof & Proof Retrieval & Academic\\
Stacks\_Proof & \citet{Welleck2021NaturalProofsMT} & Math Statement & Proof & Proof Retrieval & Academic\\
Stein\_Proof & \citet{Welleck2021NaturalProofsMT} & Math Statement & Proof & Proof Retrieval & Academic\\
Trench\_Proof & \citet{Welleck2021NaturalProofsMT} & Math Statement & Proof & Proof Retrieval & Academic\\
ProofWiki\_Reference & \citet{Welleck2021NaturalProofsMT} & Math Statement & Reference & Reference Retrieval & Academic\\
Stacks\_Reference & \citet{Welleck2021NaturalProofsMT} & Math Statement & Reference & Reference Retrieval & Academic\\
\bottomrule
\end{tabular}
\end{table*}

\begin{table*}[ht]
\centering\small
\setlength\tabcolsep{1.5pt}
\caption{List of tasks in \textsc{Mair}.}
\label{table:list-tasks-2}
\begin{tabular}{l cccccc}
\toprule
\textbf{Task} & Reference & Query & Doc & Task Type & Domain\\
\midrule
ClinicalTrials\_2021 & \href{https://www.trec-cds.org/2021.html}{TREC CDS 2021} & Health Record & Clinical Trials & Medical IR & Medical\\
ClinicalTrials\_2022 & \href{https://www.trec-cds.org/2022.html}{TREC CDS 2022} & Health Record & Clinical Trials & Medical IR & Medical\\
ClinicalTrials\_2023 & \href{https://www.trec-cds.org/2023.html}{TREC CDS 2023} & Patient Description & Clinical Trials & Medical IR & Medical\\
Genomics-AdHoc\_2004 & \href{https://trec.nist.gov/data/genomics.html}{Genomics} & Question & Document & Medical IR & Medical\\
Genomics-AdHoc\_2005 & \href{https://trec.nist.gov/data/genomics.html}{Genomics} & Question & Document & Medical IR & Medical\\
Genomics-AdHoc\_2006 & \href{https://trec.nist.gov/data/genomics.html}{Genomics} & Question & Document & Medical IR & Medical\\
Genomics-AdHoc\_2007 & \href{https://trec.nist.gov/data/genomics.html}{Genomics} & Question & Entity \& Passages & Medical IR & Medical\\
\bottomrule
\end{tabular}
\end{table*}

\begin{table*}[h]
\centering\tiny
\setlength\tabcolsep{2pt}
\caption{Results (nDCG@10) on \textsc{Mair} tasks. The last group represents the re-ranking models, which all re-rank the top 100 results of \code{text-embedding-3-small}.}
\label{table:all-results-1}
\begin{tabular}{l cc cc cc cc cc cc cc cc cc cc}

\toprule
\textbf{Model} & \multicolumn{2}{c}{\textbf{Math}} & \multicolumn{2}{c}{\textbf{PW\_Proof}} & \multicolumn{2}{c}{\textbf{PW\_Ref}} & \multicolumn{2}{c}{\textbf{Stacks\_Proof}} & \multicolumn{2}{c}{\textbf{Stacks\_Ref}} & \multicolumn{2}{c}{\textbf{Stein\_Proof}} & \multicolumn{2}{c}{\textbf{Stein\_Ref}} & \multicolumn{2}{c}{\textbf{Trench\_Proof}} & \multicolumn{2}{c}{\textbf{Trench\_Ref}} & \multicolumn{2}{c}{\textbf{TAD}} \\
\midrule
Instruction? & \ding{55} & \ding{51} & \ding{55} & \ding{51} & \ding{55} & \ding{51} & \ding{55} & \ding{51} & \ding{55} & \ding{51} & \ding{55} & \ding{51} & \ding{55} & \ding{51} & \ding{55} & \ding{51} & \ding{55} & \ding{51} & \ding{55} & \ding{51} \\
\midrule
BM25 & 50.82 & - & 49.71 & - & 12.23 & - & 28.74 & - & 24.18 & - & 59.01 & - & 23.47 & - & 63.16 & - & 22.13 & - & 26.16 & - \\
\midrule

\code{contriever-msmarco} & 47.67 & - & 38.04 & - & 11.76 & - & 13.19 & - & 20.31 & - & 50.51 & - & 20.76 & - & 43.16 & - & 19.56 & - & 18.71 & - \\
\code{gtr-t5-base} & 60.22 & - & 33.65 & - & 12.11 & - & 16.33 & - & 16.68 & - & 51.58 & - & 23.73 & - & 38.12 & - & 21.83 & - & 20.89 & - \\
\code{gtr-t5-large} & 70.21 & - & 44.40 & - & 11.96 & - & 20.69 & - & 17.80 & - & 60.47 & - & 26.36 & - & 46.13 & - & 23.23 & - & 22.17 & - \\
\midrule

\code{all-MiniLM-L6-v2} & 79.14 & 76.00 & 51.29 & 44.79 & 12.47 & 12.39 & 38.39 & 31.50 & 20.32 & 21.14 & 67.69 & 63.22 & 33.45 & 32.96 & 63.62 & 62.70 & 25.59 & 25.04 & 25.79 & 21.39 \\
\code{e5-small-v2} & 72.35 & 68.75 & 42.48 & 37.58 & 10.50 & 10.21 & 26.08 & 21.90 & 21.66 & 19.48 & 63.65 & 54.20 & 25.01 & 25.75 & 56.85 & 48.41 & 21.20 & 18.79 & 24.59 & 20.20 \\
\bottomrule
\end{tabular}
\end{table*}

\begin{table*}[h]
\centering\tiny
\setlength\tabcolsep{2pt}
\caption{Results (nDCG@10) on \textsc{Mair} tasks. The last group represents the re-ranking models, which all re-rank the top 100 results of \code{text-embedding-3-small}.}
\label{table:all-results-2}
\begin{tabular}{l cc cc cc cc cc cc cc cc cc cc}

\toprule
\textbf{Model} & \multicolumn{2}{c}{\textbf{ELI5}} & \multicolumn{2}{c}{\textbf{Fever}} & \multicolumn{2}{c}{\textbf{TREx}} & \multicolumn{2}{c}{\textbf{WnCw}} & \multicolumn{2}{c}{\textbf{WnWi}} & \multicolumn{2}{c}{\textbf{WoW}} & \multicolumn{2}{c}{\textbf{zsRE}} & \multicolumn{2}{c}{\textbf{AILA-Case}} & \multicolumn{2}{c}{\textbf{AILA-Statutes}} & \multicolumn{2}{c}{\textbf{BSARD}} \\
\midrule
Instruction? & \ding{55} & \ding{51} & \ding{55} & \ding{51} & \ding{55} & \ding{51} & \ding{55} & \ding{51} & \ding{55} & \ding{51} & \ding{55} & \ding{51} & \ding{55} & \ding{51} & \ding{55} & \ding{51} & \ding{55} & \ding{51} & \ding{55} & \ding{51} \\
\midrule
BM25 & 16.22 & - & 58.37 & - & 70.80 & - & 6.24 & - & 9.11 & - & 45.67 & - & 79.84 & - & 13.20 & - & 11.03 & - & 14.17 & - \\
\midrule

\code{contriever-msmarco} & 45.91 & - & 90.62 & - & 84.10 & - & 7.44 & - & 9.31 & - & 58.93 & - & 96.52 & - & 10.27 & - & 9.10 & - & 10.89 & - \\
\code{gtr-t5-base} & 41.35 & - & 83.23 & - & 70.04 & - & 9.28 & - & 5.93 & - & 57.99 & - & 89.39 & - & 4.69 & - & 9.84 & - & 14.22 & - \\
\code{gtr-t5-large} & 41.28 & - & 86.75 & - & 77.34 & - & 9.71 & - & 6.13 & - & 58.62 & - & 91.97 & - & 4.83 & - & 12.96 & - & 18.58 & - \\
\midrule

\code{all-MiniLM-L6-v2} & 43.27 & 27.33 & 74.50 & 44.85 & 71.75 & 0.43 & 7.86 & 6.20 & 6.66 & 7.02 & 54.75 & 50.82 & 84.69 & 0.00 & 6.54 & 5.76 & 11.71 & 7.24 & 8.04 & 1.17 \\
\code{e5-small-v2} & 43.96 & 42.91 & 88.16 & 83.88 & 86.31 & 13.87 & 9.49 & 7.91 & 10.03 & 11.71 & 59.22 & 61.69 & 98.50 & 10.54 & 7.60 & 6.43 & 10.41 & 7.52 & 10.10 & 7.84 \\
\bottomrule
\end{tabular}
\end{table*}

\begin{table*}[h]
\centering\tiny
\setlength\tabcolsep{2pt}
\caption{Results (nDCG@10) on \textsc{Mair} tasks. The last group represents the re-ranking models, which all re-rank the top 100 results of \code{text-embedding-3-small}.}
\label{table:all-results-3}
\begin{tabular}{l cc cc cc cc cc cc cc cc cc cc}

\toprule
\textbf{Model} & \multicolumn{2}{c}{\textbf{Monant}} & \multicolumn{2}{c}{\textbf{NTCIR}} & \multicolumn{2}{c}{\textbf{PointRec}} & \multicolumn{2}{c}{\textbf{ProCIS-Dialog}} & \multicolumn{2}{c}{\textbf{ProCIS-Turn}} & \multicolumn{2}{c}{\textbf{QuanTemp}} & \multicolumn{2}{c}{\textbf{TableSearch}} & \multicolumn{2}{c}{\textbf{CARE}} & \multicolumn{2}{c}{\textbf{MISeD}} & \multicolumn{2}{c}{\textbf{SParC}} \\
\midrule
Instruction? & \ding{55} & \ding{51} & \ding{55} & \ding{51} & \ding{55} & \ding{51} & \ding{55} & \ding{51} & \ding{55} & \ding{51} & \ding{55} & \ding{51} & \ding{55} & \ding{51} & \ding{55} & \ding{51} & \ding{55} & \ding{51} & \ding{55} & \ding{51} \\
\midrule
BM25 & 72.51 & - & 25.28 & - & 21.08 & - & 27.73 & - & 20.62 & - & 91.82 & - & 11.31 & - & 73.55 & - & 60.31 & - & 47.28 & - \\
\midrule

\code{contriever-msmarco} & 72.95 & - & 16.66 & - & 43.99 & - & 37.33 & - & 26.20 & - & 87.67 & - & 13.23 & - & 51.98 & - & 63.61 & - & 75.03 & - \\
\code{gtr-t5-base} & 75.99 & - & 12.06 & - & 35.61 & - & 47.47 & - & 33.00 & - & 87.25 & - & 12.12 & - & 35.47 & - & 55.83 & - & 26.55 & - \\
\code{gtr-t5-large} & 75.27 & - & 12.07 & - & 43.43 & - & 50.13 & - & 36.37 & - & 87.30 & - & 14.05 & - & 33.51 & - & 60.80 & - & 51.90 & - \\
\midrule

\code{all-MiniLM-L6-v2} & 74.20 & 70.12 & 15.06 & 10.19 & 27.58 & 17.97 & 41.03 & 19.41 & 32.02 & 19.35 & 79.52 & 74.10 & 12.01 & 7.27 & 41.09 & 14.63 & 57.14 & 53.14 & 62.26 & 18.16 \\
\code{e5-small-v2} & 77.68 & 75.59 & 12.88 & 8.67 & 45.11 & 42.47 & 26.69 & 7.31 & 22.28 & 11.06 & 88.65 & 88.39 & 9.97 & 8.46 & 57.97 & 48.39 & 54.83 & 51.81 & 66.65 & 32.80 \\
\bottomrule
\end{tabular}
\end{table*}

\begin{table*}[h]
\centering\tiny
\setlength\tabcolsep{2pt}
\caption{Results (nDCG@10) on \textsc{Mair} tasks. The last group represents the re-ranking models, which all re-rank the top 100 results of \code{text-embedding-3-small}.}
\label{table:all-results-4}
\begin{tabular}{l cc cc cc cc cc cc cc cc cc cc}

\toprule
\textbf{Model} & \multicolumn{2}{c}{\textbf{CliniDS\_16}} & \multicolumn{2}{c}{\textbf{CliniTrial\_21}} & \multicolumn{2}{c}{\textbf{CliniTrial\_22}} & \multicolumn{2}{c}{\textbf{CliniTrial\_23}} & \multicolumn{2}{c}{\textbf{DD\_15}} & \multicolumn{2}{c}{\textbf{DD\_16}} & \multicolumn{2}{c}{\textbf{DD\_17}} & \multicolumn{2}{c}{\textbf{FairRank\_20}} & \multicolumn{2}{c}{\textbf{FairRank\_21}} & \multicolumn{2}{c}{\textbf{FairRank\_22}} \\
\midrule
Instruction? & \ding{55} & \ding{51} & \ding{55} & \ding{51} & \ding{55} & \ding{51} & \ding{55} & \ding{51} & \ding{55} & \ding{51} & \ding{55} & \ding{51} & \ding{55} & \ding{51} & \ding{55} & \ding{51} & \ding{55} & \ding{51} & \ding{55} & \ding{51} \\
\midrule
BM25 & 20.79 & - & 22.95 & - & 16.70 & - & 44.86 & - & 35.31 & - & 29.33 & - & 27.22 & - & 37.19 & - & 60.13 & - & 81.42 & - \\
\midrule

\code{contriever-msmarco} & 6.71 & - & 22.90 & - & 21.59 & - & 27.10 & - & 35.73 & - & 28.42 & - & 35.98 & - & 40.43 & - & 52.85 & - & 80.39 & - \\
\code{gtr-t5-base} & 11.68 & - & 21.27 & - & 18.00 & - & 23.49 & - & 46.36 & - & 33.66 & - & 32.17 & - & 39.54 & - & 43.81 & - & 81.19 & - \\
\code{gtr-t5-large} & 12.26 & - & 24.10 & - & 23.78 & - & 25.27 & - & 45.33 & - & 34.99 & - & 31.65 & - & 37.01 & - & 43.22 & - & 80.00 & - \\
\midrule

\code{all-MiniLM-L6-v2} & 11.95 & 9.33 & 25.25 & 25.12 & 22.06 & 19.64 & 22.19 & 5.24 & 34.98 & 24.87 & 28.31 & 18.89 & 34.13 & 32.33 & 38.00 & 9.78 & 51.68 & 27.66 & 74.17 & 24.82 \\
\code{e5-small-v2} & 5.57 & 5.47 & 18.05 & 12.79 & 16.94 & 11.57 & 22.00 & 14.68 & 38.19 & 30.02 & 34.35 & 28.62 & 31.17 & 29.74 & 42.86 & 28.54 & 41.17 & 25.82 & 67.21 & 27.60 \\
\bottomrule
\end{tabular}
\end{table*}

\begin{table*}[h]
\centering\tiny
\setlength\tabcolsep{2pt}
\caption{Results (nDCG@10) on \textsc{Mair} tasks. The last group represents the re-ranking models, which all re-rank the top 100 results of \code{text-embedding-3-small}.}
\label{table:all-results-5}
\begin{tabular}{l cc cc cc cc cc cc cc cc cc cc}

\toprule
\textbf{Model} & \multicolumn{2}{c}{\textbf{Core17}} & \multicolumn{2}{c}{\textbf{News21}} & \multicolumn{2}{c}{\textbf{Robust04}} & \multicolumn{2}{c}{\textbf{InstructIR}} & \multicolumn{2}{c}{\textbf{NevIR}} & \multicolumn{2}{c}{\textbf{IFEval}} \\
\midrule
Instruction? & \ding{55} & \ding{51} & \ding{55} & \ding{51} & \ding{55} & \ding{51} & \ding{55} & \ding{51} & \ding{55} & \ding{51} & \ding{55} & \ding{51} \\
\midrule
BM25 & 11.51 & - & 20.79 & - & 18.57 & - & 36.31 & - & 66.76 & - & 28.28 & - \\
\midrule

\code{contriever-msmarco} & 18.18 & - & 27.33 & - & 26.64 & - & 50.45 & - & 62.97 & - & 25.51 & - \\
\code{gtr-t5-base} & 22.73 & - & 22.49 & - & 26.90 & - & 49.04 & - & 63.14 & - & 22.59 & - \\
\code{gtr-t5-large} & 22.80 & - & 26.62 & - & 29.07 & - & 50.87 & - & 59.95 & - & 23.46 & - \\
\midrule

\code{all-MiniLM-L6-v2} & 20.46 & 17.50 & 26.86 & 25.76 & 25.96 & 23.92 & 48.28 & 74.06 & 57.43 & 41.58 & 26.06 & 37.39 \\
\code{e5-small-v2} & 22.84 & 20.45 & 28.35 & 24.02 & 23.85 & 19.90 & 49.82 & 80.25 & 62.58 & 60.99 & 21.75 & 36.96 \\
\bottomrule
\end{tabular}
\end{table*}

\begin{table*}[h]
\centering\tiny
\setlength\tabcolsep{2pt}
\caption{Results of each sub task in SWE-Bench, CQADupStack, and IFEval.}
\label{table:all-sub-results-1}
\begin{tabular}{l cc cc cc cc cc cc cc cc cc cc}

\toprule
\textbf{Task} & \multicolumn{20}{c}{SWE-Bench}
\\

\textbf{Sub-Task} & \multicolumn{2}{c}{\textbf{astropy}} & \multicolumn{2}{c}{\textbf{django}} & \multicolumn{2}{c}{\textbf{matplotlib}} & \multicolumn{2}{c}{\textbf{mwaskom}} & \multicolumn{2}{c}{\textbf{pallets}} & \multicolumn{2}{c}{\textbf{psf}} & \multicolumn{2}{c}{\textbf{pydata}} & \multicolumn{2}{c}{\textbf{pylint-dev}} & \multicolumn{2}{c}{\textbf{scikit-learn}} & \multicolumn{2}{c}{\textbf{sphinx-doc}} \\
\midrule
Instruction? & \ding{55} & \ding{51} & \ding{55} & \ding{51} & \ding{55} & \ding{51} & \ding{55} & \ding{51} & \ding{55} & \ding{51} & \ding{55} & \ding{51} & \ding{55} & \ding{51} & \ding{55} & \ding{51} & \ding{55} & \ding{51} & \ding{55} & \ding{51} \\
\midrule
BM25 & 41.08 & - & 69.01 & - & 34.98 & - & 76.95 & - & 74.11 & - & 60.27 & - & 58.86 & - & 55.51 & - & 51.68 & - & 48.50 & - \\
\midrule

\code{contriever-msmarco} & 36.66 & - & 55.60 & - & 19.46 & - & 24.45 & - & 51.04 & - & 43.86 & - & 30.10 & - & 29.30 & - & 37.73 & - & 13.90 & - \\
\code{gtr-t5-base} & 48.85 & - & 64.13 & - & 16.13 & - & 25.87 & - & 56.41 & - & 65.62 & - & 33.72 & - & 28.46 & - & 37.54 & - & 50.13 & - \\
\code{gtr-t5-large} & 52.69 & - & 57.08 & - & 16.41 & - & 31.84 & - & 60.92 & - & 55.43 & - & 35.79 & - & 36.10 & - & 42.01 & - & 39.22 & - \\
\midrule

\code{all-MiniLM-L6-v2} & 36.33 & 33.47 & 61.65 & 53.21 & 17.90 & 17.69 & 21.99 & 27.21 & 62.14 & 58.69 & 35.26 & 36.33 & 31.23 & 26.02 & 27.47 & 26.77 & 41.26 & 38.26 & 30.12 & 16.34 \\
\bottomrule

\end{tabular}
\end{table*}

\begin{table*}[h]
\centering\tiny
\setlength\tabcolsep{2pt}
\caption{Results of each sub task in SWE-Bench, CQADupStack, and IFEval.}
\label{table:all-sub-results-2}
\begin{tabular}{l cc cc cc cc cc cc cc cc cc cc}

\toprule
\textbf{Task} & \multicolumn{6}{c|}{CQADupStack} & \multicolumn{14}{|c}{IFEval}\\
\textbf{Sub-Task} & \multicolumn{2}{c}{\textbf{Unix}} & \multicolumn{2}{c}{\textbf{WebMasters}} & \multicolumn{2}{c}{\textbf{Wordpress}} & \multicolumn{2}{c}{\textbf{format}} & \multicolumn{2}{c}{\textbf{keywords}} & \multicolumn{2}{c}{\textbf{punctuation}} & \multicolumn{2}{c}{\textbf{change\_case}} & \multicolumn{2}{c}{\textbf{length\_constraints}} & \multicolumn{2}{c}{\textbf{combination}} & \multicolumn{2}{c}{\textbf{content}} \\
\midrule
Instruction? & \ding{55} & \ding{51} & \ding{55} & \ding{51} & \ding{55} & \ding{51} & \ding{55} & \ding{51} & \ding{55} & \ding{51} & \ding{55} & \ding{51} & \ding{55} & \ding{51} & \ding{55} & \ding{51} & \ding{55} & \ding{51} & \ding{55} & \ding{51} \\
\midrule
BM25 & 32.80 & - & 31.60 & - & 16.31 & - & 26.81 & - & 26.67 & - & 42.69 & - & 42.85 & - & 20.91 & - & 33.11 & - & 34.10 & - \\
\midrule

\code{contriever-msmarco} & 46.18 & - & 19.83 & - & 25.00 & - & 19.87 & - & 22.40 & - & 29.02 & - & 39.50 & - & 22.20 & - & 24.50 & - & 32.82 & - \\
\code{gtr-t5-base} & 41.33 & - & 24.31 & - & 10.00 & - & 18.85 & - & 21.27 & - & 32.62 & - & 22.13 & - & 19.07 & - & 33.82 & - & 31.01 & - \\
\code{gtr-t5-large} & 47.52 & - & 18.72 & - & 26.57 & - & 19.33 & - & 17.93 & - & 34.26 & - & 25.36 & - & 21.29 & - & 32.31 & - & 31.73 & - \\
\midrule

\code{all-MiniLM-L6-v2} & 55.63 & 33.63 & 32.50 & 30.00 & 9.20 & 6.31 & 22.42 & 40.41 & 19.33 & 57.56 & 30.13 & 33.27 & 39.31 & 27.34 & 21.75 & 24.56 & 26.03 & 33.95 & 32.69 & 42.10 \\
\bottomrule
\end{tabular}
\end{table*}

\begin{table*}[h]
\centering\tiny
\setlength\tabcolsep{2pt}
\caption{Results of each sub task in SWE-Bench, CQADupStack, and IFEval.}
\label{table:all-sub-results-3}
\begin{tabular}{l cc cc cc cc cc cc cc cc cc cc}

\toprule
\textbf{Task} &  \multicolumn{2}{c}{IFEval}\\
\textbf{Sub-Task} & \multicolumn{2}{c}{\textbf{startend}} \\
\midrule
Instruction? & \ding{55} & \ding{51} \\
\midrule
BM25 & 14.55 & - \\
\midrule

\code{contriever-msmarco} & 26.57 & - \\
\code{gtr-t5-base} & 20.55 & - \\
\code{gtr-t5-large} & 22.68 & - \\
\midrule

\code{all-MiniLM-L6-v2} & 26.62 & 40.72 \\
\code{e5-small-v2} & 18.86 & 43.20 \\
\code{e5-base-v2} & 27.78 & 59.70 \\
\code{e5-large-v2} & 14.72 & 48.01 \\
\code{gte-base-en-v1.5} & 34.70 & 62.87 \\
\code{gte-large-en-v1.5} & 31.31 & 61.94 \\
\code{bge-base-en-v1.5} & 27.17 & 33.74 \\
\code{bge-large-en-v1.5} & 23.24 & 48.42 \\
\code{text-embedding-3-small} & 14.01 & 37.18 \\
\midrule

\code{e5-mistral-7b-instruct} & 35.03 & 35.23 \\
\code{NV-Embed-v1} & 17.17 & 19.75 \\
\code{GritLM-7B} & 28.59 & 31.65 \\
\code{gte-Qwen2-1.5B-instruct} & 28.39 & 53.10 \\
\midrule

\code{monot5-base-msmarco} & 16.90 & 38.37 \\
\code{bge-reranker-v2-m3} & 7.83 & 22.97 \\
\code{bge-reranker-v2-gemma} & 11.29 & 72.85 \\
\code{jina-reranker-v2-base} & 2.29 & 7.08 \\
\code{mxbai-rerank-large-v1} & 7.27 & 16.82 \\
\code{gpt-3.5-turbo} & 7.88 & 16.35 \\
\bottomrule
\end{tabular}
\end{table*}
\begin{table*}[h]
\centering\tiny
\setlength\tabcolsep{2pt}
\caption{Results (nDCG@10) on IFEval sub task. AVG means average results.}
\label{table:if-eval}
\begin{tabular}{l cc cc cc cc cc cc cc cc cc cc}

\toprule
\textbf{Model} & \multicolumn{2}{c}{\textbf{AVG}} & \multicolumn{2}{c}{\textbf{detectable\_format}} & \multicolumn{2}{c}{\textbf{keywords}} & \multicolumn{2}{c}{\textbf{punctuation}} & \multicolumn{2}{c}{\textbf{change\_case}} & \multicolumn{2}{c}{\textbf{length\_constraints}} & \multicolumn{2}{c}{\textbf{combination}} & \multicolumn{2}{c}{\textbf{detectable\_content}} & \multicolumn{2}{c}{\textbf{startend}} \\
\midrule
Instruction? & \ding{55} & \ding{51} & \ding{55} & \ding{51} & \ding{55} & \ding{51} & \ding{55} & \ding{51} & \ding{55} & \ding{51} & \ding{55} & \ding{51} & \ding{55} & \ding{51} & \ding{55} & \ding{51} & \ding{55} & \ding{51} \\
\midrule
BM25 & 28.28 & - & 26.81 & - & 26.67 & - & 42.69 & - & 42.85 & - & 20.91 & - & 33.11 & - & 34.10 & - & 14.55 & - \\
\midrule
\code{text-embedding-3-small} & 21.50 & 31.91 & 23.47 & 34.09 & 23.11 & 48.87 & 32.44 & 33.47 & 20.28 & 17.98 & 16.94 & 23.08 & 13.41 & 20.78 & 28.73 & 38.65 & 14.01 & 37.18 \\
\code{e5-mistral-7b-instruct} & 24.98 & 32.15 & 23.04 & 32.74 & 11.51 & 58.66 & 31.98 & 33.90 & 31.34 & 21.96 & 14.60 & 22.76 & 38.96 & 24.51 & 30.24 & 32.65 & 35.03 & 35.23 \\
\code{NV-Embed-v1} & 22.45 & 27.68 & 21.65 & 25.74 & 22.51 & 42.90 & 28.55 & 37.48 & 27.25 & 27.58 & 17.89 & 19.91 & 17.77 & 24.17 & 31.32 & 36.96 & 17.17 & 19.75 \\
\code{GritLM-7B} & 22.03 & 38.73 & 20.75 & 34.36 & 5.51 & 65.11 & 42.05 & 33.82 & 29.92 & 40.99 & 12.87 & 29.17 & 32.33 & 49.73 & 23.13 & 45.11 & 28.59 & 31.65 \\
\code{gte-Qwen2-1.5B-instruct} & 21.85 & 42.76 & 18.13 & 44.52 & 14.95 & 58.09 & 32.27 & 32.59 & 28.90 & 34.94 & 16.40 & 23.86 & 24.68 & 42.86 & 25.95 & 56.47 & 28.39 & 53.10 \\
\bottomrule
\end{tabular}
\end{table*}

\begin{figure*}[!t] \centering
\includegraphics[width=1\columnwidth]{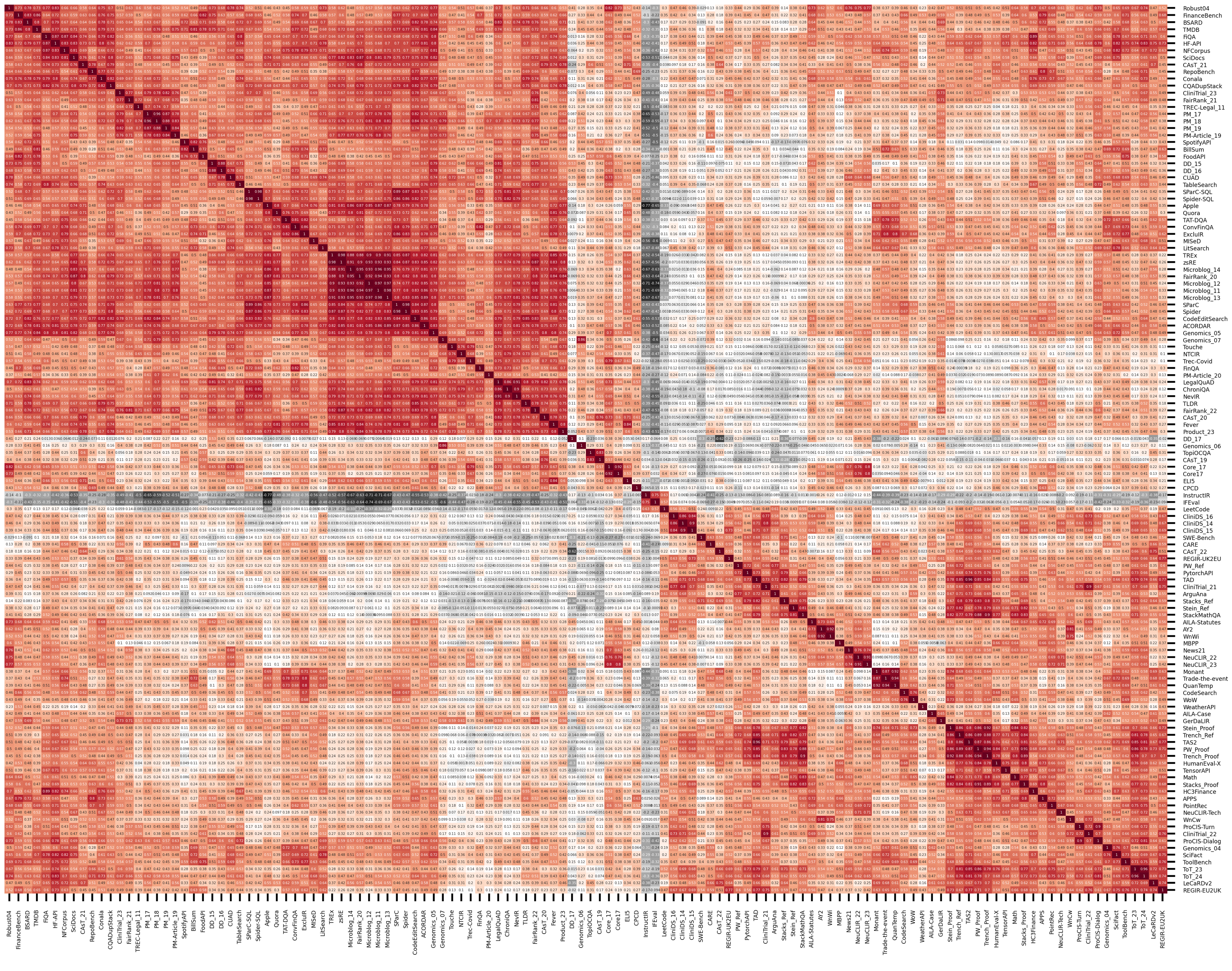} 
\caption{Task correlation heat map. The correlation between two tasks is measured by the Pearson correlation coefficient of different models' performance on these tasks. This coefficient ranges from -1 to 1. A value close to 1 indicates that models which perform well on Task A are likely to perform well on Task B. Conversely, a value close to -1 indicates that models which perform well on Task A are likely to perform poorly on Task B.}
\label{fig:heat}
\end{figure*}

\clearpage

\end{document}